\documentstyle[aps,preprint]{revtex}

\begin{document}
\draft
\title{Current profiles and AC losses of a superconducting strip with
elliptic
cross-section in perpendicular magnetic field}
\author{F. G\"om\"ory$^{1}$, R. Tebano$^{2}$, A.
Sanchez$^{3}$, E. Pardo$^{3}$, C. Navau$^{3}$, I. Husek$^{1}$, F.
Strycek$^{1}$, and P.
Kovac$^{1}$}
\address{
$^1$Institute of Electrical Engineering, Slovak Academy of
Sciences,
D\'ubravsk\'a 9,\\
842 39 Bratislava, Slovak Republic\\
$^2$Dipartimento di Scienza dei Materiali, Istituto Nazionale per
la Fisica della Materia, INFM\\
Universit\`{a} degli Studi di
Milano-Bicocca,
Via Cozzi 53, 20125 Milano, Italy\\
$^3$Grup d'Electromagnetisme,
Departament de F\'\i sica, Universitat Aut\`onoma Barcelona \\
08193 Bellaterra (Barcelona), Catalonia, Spain}

\maketitle
\begin{abstract}

The case of a hard type II superconductor in the form of strip with
elliptic
cross-section when placed in transverse magnetic field is studied.
We
approach the problem in two steps, both based on the critical-state
model. First we 
calculate numerically the penetrated current
profiles that ensure complete shielding in the interior, without
assuming an 
{\it a priori} form for the profiles. In the second step we introduce
an
analytical approximation that asumes that the current profiles are
ellipses.
Expressions linking the sample magnetization to the applied field are
derived covering the whole range of applied fields. The theoretical
predictions are tested by the comparison with experimental data for
the
imaginary part of AC susceptibility.

\end{abstract}


\section{Introduction}

The successful use of high-$T_{c}$ superconductors in the fabrication
of
conductors for high current applications requires a deep
understanding of
their response to the magnetic field and, in particular, their ac
losses
behavior. An important case to study is the magnetic response of a
tape made
from hard type II superconductor, with a cross-section of elliptic
shape, to
a magnetic field applied perpendicular to its major axis. This is the
configuration very often met in practice and includes the common case
of a
cylindrical wire in perpendicular field. There were several attempts
to
study this system, all based in the critical-state model \cite{bean}.
Within
this model framework, the application of a magnetic field in a
superconductor results in the penetration of current with a constant
density 
$J_{c}$. The current profile corresponding to a given applied field
is
distributed in such a way that it keeps the field in the internal
region
unchanged (so, in the initial magnetization curve, the interior field
is
kept zero). Once a method is found to obtain the actual shape for the
current profiles -also called flux fronts-, all the magnetic
properties such
as magnetization, ac susceptibility and ac losses can be deduced from
it.

A first approach to the problem of a superconducting tape with
elliptical
cross-section assumed that the flux fronts are ellipses with one
constant
axis (that in the direction of the applied field) and the other axis 
varying in order to match the field change \cite{Wilson,krasnov}.
However,
the assumption of the constant axis was shown to be incorrect in
computations of cylindrical wires in transverse field \cite{ashkin},
spheres
and spheroids \cite{navarro,telschow} and also thick strips
\cite{Brandt96},
where it was shown that field profiles detach from the surface.

Another important point is the actual shape of the profiles. Ashkin
\cite
{ashkin} developed a numerical technique to calculate current
penetration
profiles by forcing the field on the current penetration profile to
vanish.
Although the magnetic field in the current-free region was zero, as
required
by critical-state model, with a precision of only around 10 \%, the
current
profiles results showed a tendency towards a spindle shape (similar
to the
profiles shown in solid line in Fig. 1). This fact was further
pursued by
Kuzovlev \cite{kuzovlev} who demonstrated that, at least for the case
of a
sphere, the current profiles have indeed a spindle shape, and not a
more
smooth shape such as an ellipsoid. However, Bhagwat and Chaddah
solved the
case of a very thin elliptical tape assuming elliptical flux fronts
\cite
{BhagwatChaddah}.

Therefore, the situation on such an important system remains
unsolved, and
questions arise as what the actual shape of current profiles is
(elliptical
or spindle) and how correct and accurate the various approximations
presented up to now are. In this work, we will use a numerical
procedure
based on the critical-state model to accurately determine the flux
fronts
that shield the central region of the sample, for any applied field
and
sample aspect ratio. A key feature of our approach is that, different
from
the above mentioned models, we will not assume any {\it a priori}
shape for the flux fronts.
After briefly describing the main characteristics of the calculated
profiles, we will introduce an analytical model which
reproduces
the features of the actual profiles with enough precision for most
needs. We
will finally compare our results with experimental data measured on
an
actual superconducting tape.

\section{Numerical Results}

\label{MME}

Our numerical model is based on minimizing the magnetic energy of the
current
distribution after each applied field variation. This approach has
been
successfully applied to describe the experimental features observed
in
the
initial magnetization\cite{Alvaro1} as well as the whole
magnetization loop
and levitation force \cite{Alvaro2} of superconducting cylinders. The
details of the model can be found in \cite{Alvaro1,Alvaro2}.

Calculated profiles are shown in Fig. 1 (solid lines) for the cases
of superconducting tapes of elliptical cross-section with
$b/a$=0.1
(thin ellipse), 1 (circular wire), and 10 (long ellipse), where $a$
and $b$
are the ellipse semiaxes. The external field is applied in the
direction of
the (vertical) $b$ axis. In order to confirm the validity of our
approach,
we checked that the field in the non-penetrated region is zero with a
precision of around 0.1 \%. As a further check of the model, we have
found
that our calculations for thin elliptical tapes with the aspect ratio
smaller or equal than $b/a=0.01$ coincide within numerical accuracy
with the
analytical formulas for thin ellipses of Bhagwat and Chaddah
\cite{BhagwatChaddah}.
This
agreement is not observed for thicker samples. Another characteristic
of the
data is that they follow the known dependence proportional to $\cos
\theta $
predicted for round wires in low applied fields \cite{carr}. Finally,
it is important to remark that the results from our numerical
approach have been further confirmed by calculations based on the
analogous, although independent, approach proposed by Brandt
\cite{Brandt96}.

The calculations show several interesting features, which allow us to
answer
the unsolved questions posed above. First, all the profiles are
spindled-shaped, although when the elliptical cross-section of the
tape is
large in the direction of the applied field (case $b/a=10$, for
example) the
shape of the profiles resemble more that of an ellipse. Also, in all
cases
the profiles detach from the surface in the $b$ axis. The effect is
more
clear for thick samples than for thin ones. This means that the
classical
approach of Wilson \cite{Wilson}, which assumed no detachment, could
work
for thin samples but it is not expected to describe accurately the
situation
of a cylindrical wire in perpendicular applied field. Another
interesting
feature is that the spacing between the successive field fronts is
rather
constant for the thicker ellipses but not for the thinner ones.

\section{analytical approximation}

\label{GEA}

After having obtained the adequate numerical description of the
actual
current profiles for the superconducting tape with elliptical
cross-section,
we would like to know if these results could be interpolated by an
opportune
analytical expression. Here the motivation stems from the fact that
analytical approaches have important practical advantages. Therefore
we worked out an analytical model that, in spite of its simplicity, 
reproduces the magnetic results of the numerical model with
sufficient accuracy.

The model is based on the assumption that flux fronts are ellipses
with
semiaxes $a_{0}\varepsilon $ and $b_{0}\varepsilon ^{1/n}$, where
$a_{0}$
and $b_{0}$ are the semiaxes defining the strip's cross-section (see
Fig.
2). Each ellipse is characterised by the independent variable
$\varepsilon $%
, with the values ranging from 1 (flux front coinciding with the
surface) to
0 (flux front collapsed to the centre of the ellipse). To generalise
the
previous approaches, an additional parameter characterising the shape
of the
flux front, $n,$was introduced . Indeed, the limit of $n\to \infty $
corresponds to the constant axis model used for a round
wire\cite{Wilson},
while $n=2$ would reproduce the assumptions in \cite{BhagwatChaddah}.
We
will see that the approximation of flux fronts for tapes with
different
aspect ratios will require to adjust $n$ accordingly.

In the critical state approach we utilise here, the current
distribution in
an elliptic strip with aspect ratio $\beta =b_{0}/a_{0}$ is expressed
in
polar coordinates as (Fig. 2) 
\begin{eqnarray}  \label{current}
j(\varepsilon, r, \theta)=\left\{ 
\begin{array}{ll}
+j_c & \text{for} \; r_\varepsilon<r<r_1 \bigcap -\pi/2<\theta<\pi/2
\\ 
0 & \text{for} \; r<r_\varepsilon \bigcup r>r_1 \\ 
-j_c & \text{for}\; r_\varepsilon<r<r_1 \bigcap \pi/2<\theta<3\pi/2
\end{array}
\right.
\end{eqnarray}
where the outer shape of the strip is defined by
$r_1(\theta)=b_0/\sqrt{
\beta^2{\rm cos}^2(\theta)+{\rm sin}^2(\theta)}$ and the flux front
is given
as $r_\varepsilon(\theta)=\varepsilon b_0/\sqrt{\beta^2{\rm cos}
^2(\theta)+\varepsilon^{2-2/n}{\rm sin}^2(\theta)}$ . The current
distribution (\ref{current}) generates in the ellipse center the
magnetic
field in the $y$-direction 
\begin{eqnarray}  \label{fieldcenter}
H_y(\varepsilon)&=&{\frac{-2j_c}{{\pi}}}\int_0^{\pi/2}{\rm
cos}\theta\left(
r_1(\theta)-r_\varepsilon(\theta)\right)d\theta  \nonumber \\
&=&-H_p(g(1)-g(\varepsilon))
\end{eqnarray}
where $H_p=2J_cb_0/\pi$ is the penetration field for a round wire of
radius $b_0$, and $g(x)$ is the auxiliary function 
\begin{equation}  \label{g}
g(x)={\frac{x^{1/n}{\rm
arctan}\sqrt{{\frac{\beta^2}{{x^{2-2/n}}}}-1}} {{
\sqrt{{\frac{\beta^2}{{x^ {2-2/n}}}}-1}}}},
\end{equation}
which depends also on parameters $\beta$ and $n$ \cite{footnote}. An
important limit for any $\beta$ and $n$ is $g(0) = 0$. The magnetic
moment
per unit length of the strip with the current distribution
(\ref{current})
extended to length $l>>a_0,b_0$ in both $+z$ and $-z$ directions is
calculated as 
\begin{eqnarray}
{\frac{m(\varepsilon)}{{l}}}&=&-4j_c\int_0^{a_0}xb_0\sqrt{1-{\frac{x^
2 }{{\
a_0^2}}}} dx \\
-4j_c \int_0^{\varepsilon a_0}&x& \varepsilon^{1/n}
b_0\sqrt{1-{\frac{x^2}{{
\varepsilon^2 a_0^2}}}} dx ={\frac{-4j_c b_0 a_0^3}{{3}}}\left(
1-\varepsilon^{2+1/n} \right)  \nonumber
\end{eqnarray}
Then, after dividing by the tape cross-section we obtain for the
magnetization 
\begin{equation}  \label{mag}
M(\varepsilon)={\frac{m(\varepsilon)}{{\pi a_0 b_0
l}}}={\frac{-2H_p}{{
3\beta }}}\left( 1-\varepsilon^{2+1/n} \right)
\end{equation}
In the case when the whole section of the strip is saturated with the
critical current density, $\varepsilon=0$ and one can find the
penetration
field $H_s=-H_y(0)=H_p g(1)$ and the saturation magnetization
$M_s=\vert
M(0)\vert=2H_p/3\beta$.

Let us now describe the magnetization of the elliptical tape. For a
zero-field cooled sample, in the virgin stage of magnetization, the
distribution of currents, which is determined by the parameter
$\varepsilon_0
$, should be such that it shields a field $H_0^{shi}$ equal to the
applied
field $H_a$. From this condition and Eq. (\ref{fieldcenter}) the
implicit
relation linking $H_a$ with $\varepsilon_0$ in the initial
magnetization
curve is found as 
\begin{eqnarray}
H_i^{shi}&=&H_p[g(1)-g(\varepsilon_i)] \hspace{1cm} H_i^{shi}< H_s
\nonumber
\\
\varepsilon_i&=&0 \hspace{1cm} H_i^{shi}\geq H_s  \label{hshi}
\end{eqnarray}
with $H_a=H_i^{shi}$ and $i=0$. The initial magnetization $M_0$, as a
function of $\varepsilon_0$, is given by Eq. (\ref{mag}) as 
\begin{equation}
M_0(\varepsilon_0)=-\frac{2H_p}{3\beta} \left( 1-
\varepsilon_0^{2+\frac{1}{n}} \right)  \label{eini}
\end{equation}

We now analyze the dynamics of flux penetration at the AC field
$H_{a}=H_m{\rm cos}(\omega t)$, where the maximum field $H_m$
corresponds to
a $\varepsilon_m$ defined as in Eq. (\ref{hshi}) for $i=m$ and
$H_m^{shi}=H_m
$, so that $g(\varepsilon _{m})=g(1)-H_{m}/H_{p}$ if $H_{m}< H_s$,
and $\varepsilon_m=0$ otherwise.

The amplitude susceptibility, defined as the ratio of the
magnetization and
the applied field at $H_{a}=H_{m}$ \cite{GoSUST} is 
\begin{equation}
\chi _{a}={\frac{M_{0}(\varepsilon _{m})}{{H_{a}}(\varepsilon
_{m})}}=-{\frac{2}{{3\beta }}}{\frac{1-\varepsilon
_{m}^{2+1/n}}{{g(1)-g(\varepsilon
_{m})}}}
\end{equation}
At very small amplitudes, the strip is shielded by currents that
flow only in a thin surface shell and $\varepsilon _{m}\to 1$ . The
absolute value of the amplitude susceptibility in this limit, denoted
$\chi_{0}$ 
\cite{FabbPRB} is then 
\begin{equation}
\chi _{0}={\lim }_{\varepsilon _{m}\to 1}|\chi
_{a}|={\frac{2}{{3\beta }}}{
\frac{(2n+1)(\beta ^{2}-1)}{{(n\beta ^{2}-1)g(1)-n+1}}}  \label{chi0}
\end{equation}
The shielding at low penetrations is a
current distribution for which an analytic solution is known
\cite{Osborn}.
>From this solution it follows that the value of $\chi _{0}$ is 
$\chi_{0,
{\rm analytic}}=1+1/\beta $. The correspondence of the value of
$\chi_{0}$
from (\ref{chi0}) with that calculated according to the analytic
solution
could be then used as a criterion in finding the optimum value of
parameter $n$ in our model for each value of the parameter $\beta$ as 
\begin{equation}
n_{{\rm opt}}={\frac{2\beta +3g(1)-5}{{3\beta ^{2}g(1)-4\beta +1}}}.
\label{nopt}
\end{equation}

When the applied field is descending (from $H_{m}$ to $-H_{m}$)
currents in
the opposite sense begin to enter from surface to inside the tape.
Whereas
the already present currents are kept frozen the new entering
currents
should shield a field
$H_{1}^{shi}=(H_{m}-H_{a})/2$\cite{bean,ClemSanchez}.
The current profile which shields this field is defined by a
$\varepsilon
_{1}$ similarly as in Eq. (\ref{hshi}) for $i=1$. So, the
magnetization for
a given applied field $H_{a}$ (a given $H_{1}^{shi}$), considering
both new
and frozen currents, is 
\begin{eqnarray}  \label{m1b}
M_{1}(\varepsilon _{1})&=&M_{0}(\varepsilon _{m})-2M_{0}(\varepsilon
_{1}) \\
&=&{\ \frac{2H_{p}}{{3\beta }}}(\varepsilon
_{m}^{2+1/n}+1-2\varepsilon
_{1}^{2+1/n}),  \nonumber
\end{eqnarray}
where $\varepsilon _{1}$ is related with the magnetic field as $
g(\varepsilon _{1})=g(1)-(H_{m}- H_{a})/2H_{p}$ if $H_{a}> H^{*}$,
and $\varepsilon_1=0$ otherwise, and $H^{*}=H_{m}-2H_{s}$ is the
penetration
field of the reverse supercurrents\cite{chengoldfarb}. Notice that if
$H_{m}<H_{s}$, during the reversal stage, reverse currents will never
surpass
the initial ones, $\varepsilon _{1}\geq \varepsilon _{m}>0$. 
Analogous magnetization expressions can be easily derived for the
ascending
part of the cycle (from $-H_{m}$ to $H_{m}$).

We can now proceed with the calculation of $\chi ^{\prime \prime }$,
the
imaginary part of the complex AC susceptibility defined as
\cite{spectra} 
\begin{equation}
\chi ^{\prime \prime }={\frac{2}{{\pi H_{m}^{2}}}}
\int_{-H_{m}}^{H_{m}}M_{1}(H_{a})dH_{a}  \label{chi}
\end{equation}
This quantity allows to calculate the AC loss \cite{FabbPRB}. The
integral (\ref{chi}) is easily determined in a numerical way, by
inserting the
corresponding expressions for the magnetization $M_{1}(H_{a})$ for
each
stage.

The usefulness of the analytical model described above is that it
provides a
very convenient approximation to the numerical calculations. In Fig.
1 we
show (dashed lines) the current profiles corresponding to the
numerical
ones. The agreement between both is very good except in the region
closest
to the $b$ axis. However, this region is the one that contributes
less to
the magnetic moment of the sample. Significance of neglecting the
difference
between the numerical profiles and the analytical approximation is at
best
evaluated by comparing the results for the magnetization and AC
susceptibility. We found for these quantities that the results
calculated
from the analytical model are hardly distinguishable in practice
from the
accurate numerical data.

\section{Experimental verification}

Experimental test of these models were performed on a monocore tape
from
high-$T_{c}$ superconductor Bi$_{2}$Sr$_{2}$Ca$_{2}$Cu$_{3}$O$_{10}$
in
silver matrix. The data are presented in Fig. 3, where we show $\chi
^{\prime \prime }$ as function of the amplitude of the AC field,
$H_{m}$. To
compare better the shapes of the experimental curves measured at
different
temperatures with the model proposed here, all the curves were
normalized to
meet in the maximum point of the $\chi ^{\prime \prime }$ curve. Our
theoretical results correspond to both the analytical and numerical
model
(they cannot be distinguished in the scale of the picture). It is
important
to remark that this is a zero-parameter fit, since the value of
$\beta $ is
obtained from the actual sample dimensions and $n$ is chosen from
$\beta $
after Eq. (\ref{nopt}). We see clear distinction between the data
obtained
at superimposed field with respect to those measured without DC
field.
We
explain this by the known fact that applying the DC field much larger
than
the AC field amplitude limits the actual magnetic fields to a narrow
interval on the $j_{c}(B)$ dependence, approaching the assumption of
field-independent $j_{c}$. Thus, only the curves measured with
superimposed
DC field should be compared with our model derived under the
assumption of
constant  $j_{c}$. Indeed, we see that the data registered with
superimposed
DC field coincide nicely with the curve predicted by our model in the
region
about the maximum.  Slight deviations observed at low fields could be
attributed to sample imperfections. On the other side, at large AC
amplitudes - up to 0.015 T used in our experiments - the assumption
about
constant $j_{c}$ does not hold anymore because a wide range of local
magnetic
fields could lead to quite different actual values of $j_{c}$. This,
in our
opinion, explains also the huge deviation of data measured in zero DC
field
from our prediction - one can expect under these circumstances a
significant
deformation of flux front shapes due to the dependence of critical
current
density on the magnetic field. Similar conclusion was drawn also in
another
paper tackling the same problem\cite{tenhaken}.   The narrowing of
the $\chi
^{\prime \prime }(H_{a})$ curve due to $j_{c}(B)$ was predicted for
slabs\cite{spectra}, thin films\cite{shantsev} and
cylinders\cite{brandt98}. 

\section{conclusions}

We have numerically calculated the current penetration profiles of a
strip
of elliptical cross-section in a perpendicular applied field, solving
some
open questions about the shape and properties of the profiles.
Moreover, we
have presented evidence that, for tapes with any aspect ratio,
assuming that the flux fronts are
ellipses where the axis in the field direction shrinks as the
power $1/n$ with respect to the shrinking of the perpendicular axis
is a good
approximation, so they provide good
basis for a simple calculation of such properties as magnetization
loops, susceptibilities and AC losses.

\section{Aknowledgements}

We thank MCyT project BFM2000-0001, CIRIT project
1999SGR00340, and DURSI (Generalitat de Catalunya) as well as the
Slovak Grant Agency
VEGA for financial support.

\begin{figure}[tbp]
\caption{Current profiles for strips with elliptical cross-section of
semiaxes $b/a=0.1, 1,$ and 10, corresponding to applied fields
$H_{{\rm a}}=0, 0.2, 0.4, 0.6, 0.8,$ and 1, in units of the
penetration field
$H_{s}$
(from surface inwards). The strip cross-sections have been scaled as
circles. Solid lines correspond to the numerical method (Sec.
\ref{MME}), while dashed lines to the analytical approximation (Sec.
\ref{GEA}).} 
\end{figure}

\begin{figure}[tbp]
\caption{Sketch of the cross section of the elliptical tape.}
\end{figure}

\begin{figure}[tbp]
\caption{Imaginary part of the AC susceptibility,
$\chi^{\prime\prime}$, as
function of the amplitude of the applied ac field, $H_a$. The solid
line corresponds to
theoretical
values, and symbols to experimental data obtained from the tape shown
in the inset ($\beta=0.098$), for different temperatures and dc field
values. }
\end{figure}


\begin{references}
\bibitem{footnote}  Actually, Eq. (\ref{g}) as written is valid even
for the
case $\beta <x^{1-1/n}$, if using that arctan($ix$)=$i$arctanh($x$).
\end{references}
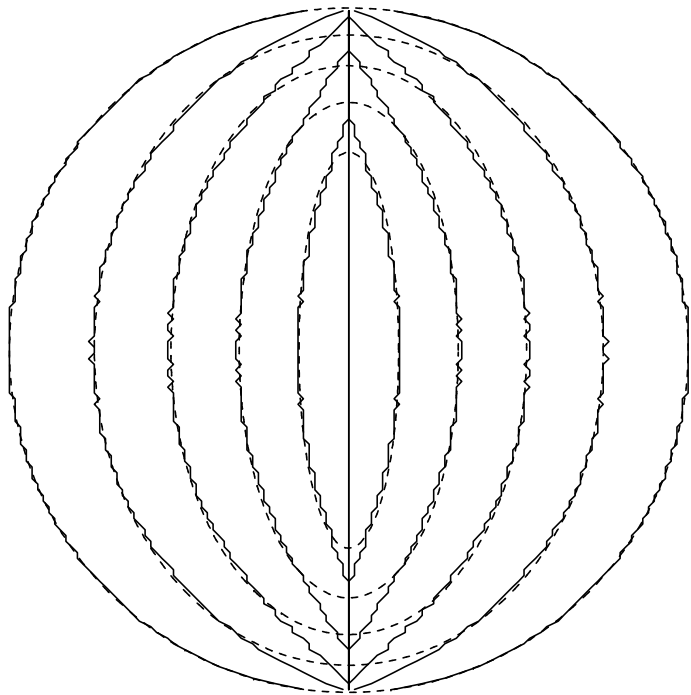
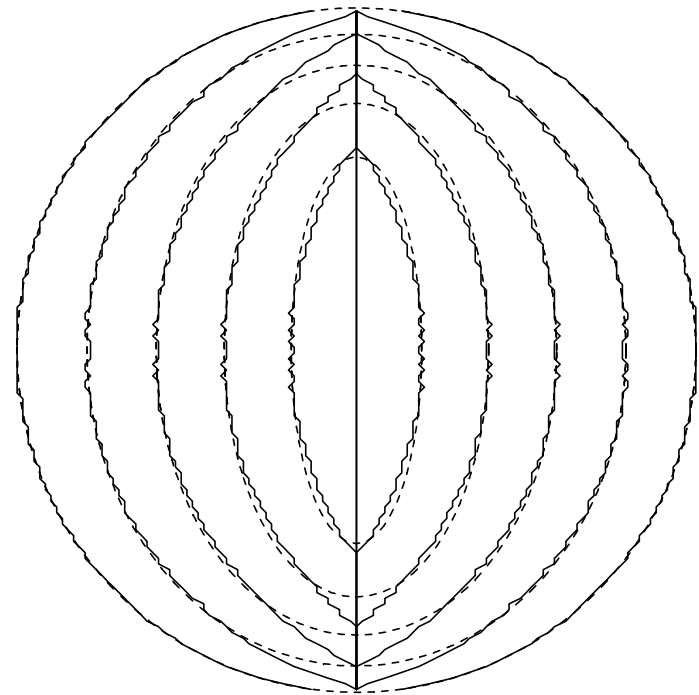
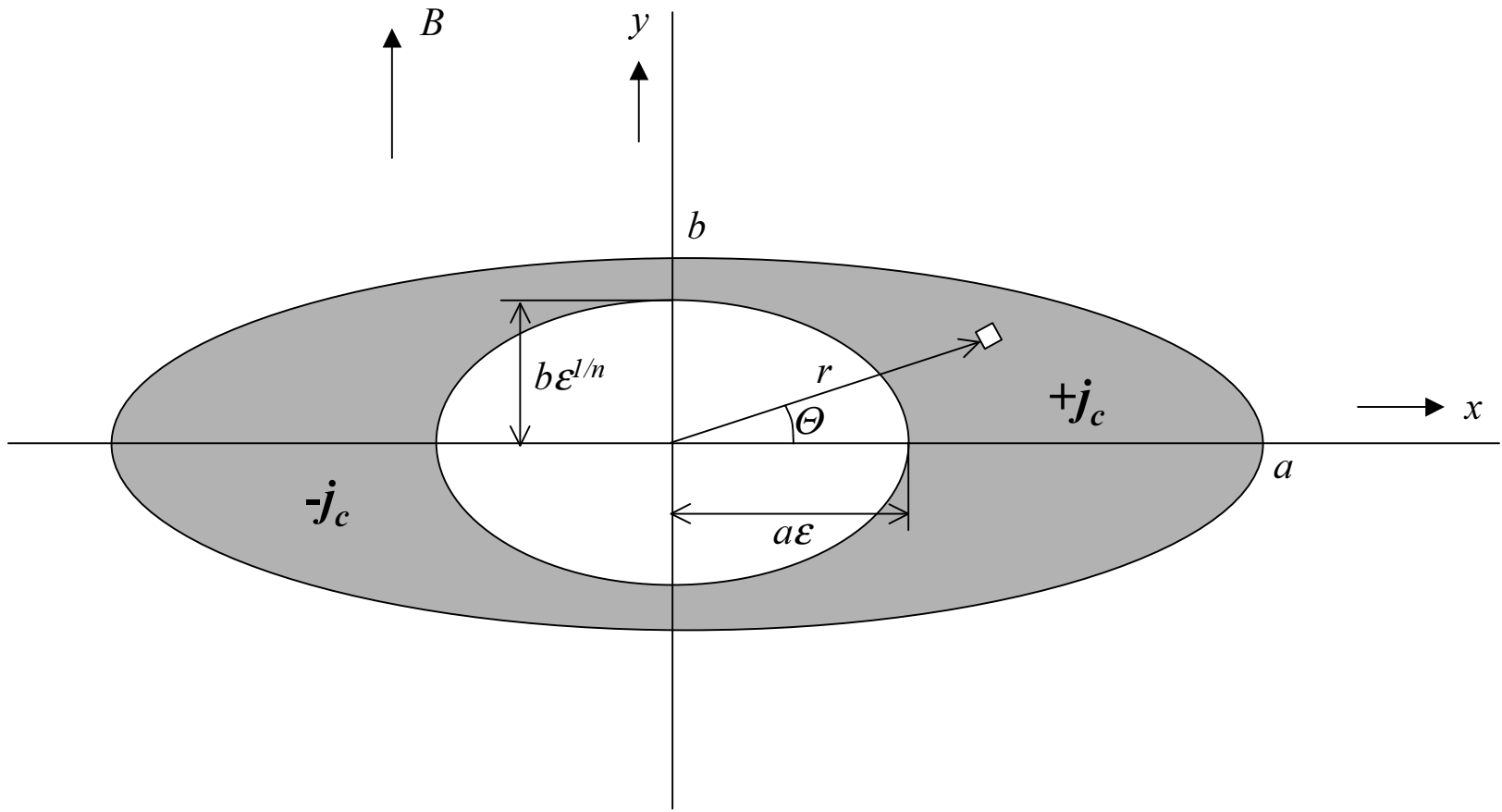
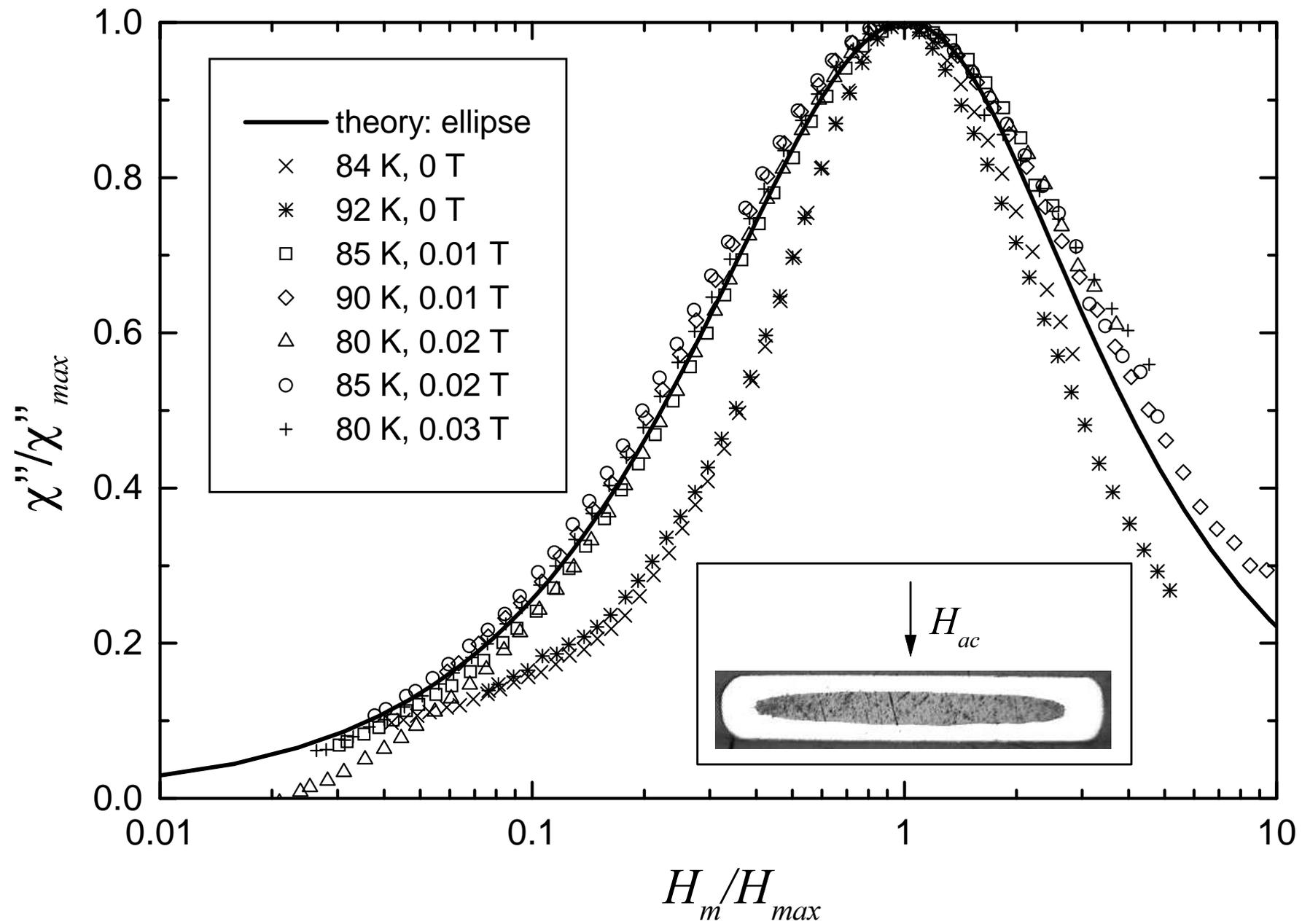
\end{document}